\documentclass{IOS-Book-Article}

\usepackage{mathptmx}
\usepackage{booktabs} 
\usepackage{subcaption}

\usepackage{graphicx}
\usepackage{float}
\usepackage[table]{xcolor} 
\usepackage{url}

\newcommand{\wrt}{{\it w.r.t. }}   
\newcommand{\eg}{\emph{e.g.}, }       
\newcommand{\ie}{\emph{i.e.}, }      
\newcommand{\etc}{\emph{etc. }}

\def\hb{\hbox to 10.7 cm{}}

\begin{document}

\pagestyle{headings}
\def\thepage{}

\begin{frontmatter}              

\title{Random Forest as a Tumour Genetic Marker Extractor}

\markboth{}{May 2019\hb}

\author[A]{\fnms{Raquel} \snm{P\'erez-Arnal}}, %
\author[A]{\fnms{Dario} \snm{Garcia-Gasulla}},
\author[A]{\fnms{David} \snm{Torrents}}, %
\author[A]{\fnms{Ferran} \snm{Par\'es}}, %
\author[A,B]{\fnms{Ulises} \snm{Cort\'es}},
\author[A,B]{\fnms{Jes\'us} \snm{Labarta}}
and
\author[A,B]{\fnms{Eduard} \snm{Ayguad\'e}}
\address[A]{Barcelona Supercomputing Center (BSC)\\ (\{raquel.perez, dario.garcia\}@bsc.es)}
\address[B]{Universitat Polit\`ecnica de Catalunya - BarcelonaTech (UPC) }

\begin{abstract}
Finding tumour genetic markers is essential to biomedicine due to their relevance for cancer detection and therapy development. In this paper, we explore a recently released dataset of chromosome rearrangements in 2,586 cancer patients, where different sorts of alterations have been detected. Using a Random Forest classifier, we evaluate the relevance of several features (some directly available in the original data, some engineered by us) related to chromosome rearrangements. This evaluation results in a set of potential tumour genetic markers, some of which are validated in the bibliography, while others are potentially novel.

\end{abstract}

\begin{keyword}
cancer research \sep chromosomal rearrangements \sep tumour genetic markers \sep Random Forest
\end{keyword}
\end{frontmatter}
\markboth{May 2019\hb}{May 2019\hb}

\section{Introduction}
Cancer is among the four current leading causes of death before the age of 70, having around 18.1 million deaths in 2018 \cite{Bray2018}. For this reason,  studying and understanding the biology of tumours constitutes  a priority in biomedicine. One of the leading research lines on this field is the study of chromosomal rearrangements in solid tumour cells. Chromosomal rearrangements (or breaks) are changes in the basic structure of a chromosome, examples of such alterations are the deletion, duplication or reordering of a subset of genes of the chromosome.

Several studies have shown that the presence of chromosomal rearrangements in tumours is often correlated with poor prognosis \cite{chromosome_aberrations_in_solid_tumours,Luijten2018}, and some of them have been identified as hallmarks of several tumour types.
This implies that the presence of some specific gene expressions or DNA changes, like chromosomal rearrangements, can be used as tumour markers  to characterize different types of cancer. Finding these markers can be useful in several ways, like predicting disease outcome or response to treatment. Some examples of chromosomal rearrangements as tumour genetic markers are mutations of chromosome 5q21 on colorectal cancer \cite{Nishisho665} or deletions on chromosome 3p on lung cancer \cite{Whang-Peng181}. 

\begin{figure}[t]
\centering
\includegraphics[width=\textwidth]{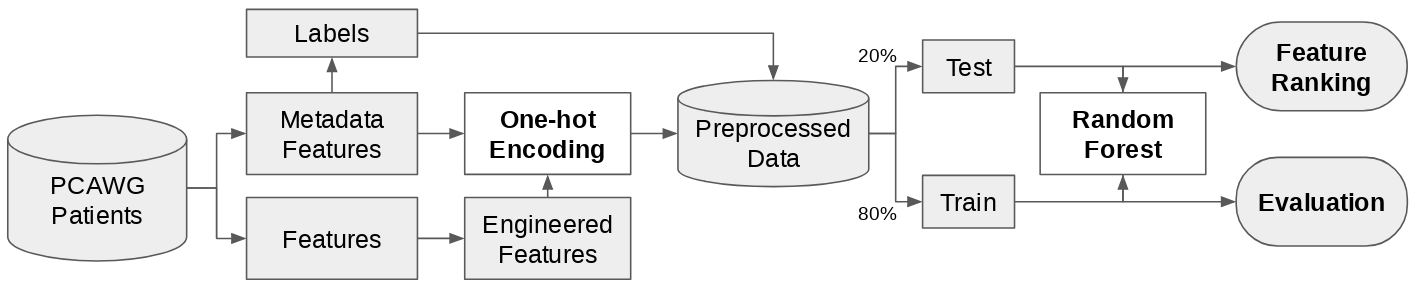}
\caption{Summary of the data pipeline. \label{fig:schema}}
\end{figure}

In this paper, we present a new methodology to find potential tumour genetic markers from a dataset of chromosomal rearrangements. This data consists of a set of files, each one of them from a patient, where we have a sequence of rearrangements, represented as triplets \textit{source base pair}, \textit{destiny base pair} and \textit{rearrangement type}. From these sequences of triplets, we engineer several features related to the chromosomal rearrangements, and then, use a Random Forest as a feature extractor. In this way, we generate a ranking of the features by their importance. The best features of the ranking are new potential genetic markers found by the methodology. In our experiments, we extract more than thirty potential markers. 
Figure \ref{fig:schema} shows the basic data pipeline used on this work.

\section{Data}\label{sec:data}
The data used in this study comes from genetic cancer data from the PanCancer Analysis of Whole Genomes (PCAWG) project \cite{Zhang2015}. The PCAWG study is an international collaboration to identify common patterns of mutation in more than 2,800 cancer whole genomes from the International Cancer Genome Consortium. Let us remark that scientific works with a primary focus on pan-cancer are under publication embargo until July 25, 2019. For this reason, there is no comparison of our results with similar approaches. This situation also guarantees the novelty of all our experiments.

Using an \textit{in-house} pipeline designed to analyse tumour genomes in a clinical and research context, we have identified breakpoints that inform of sites of genomic and chromosomal rearrangements in the PCAWG data. This pipeline identifies breakpoints using the information of reads, and paired-end reads mapping from whole-genome sequencing, using Burrows-Wheeler Aligner (BWA). These predictions have passed strict filtering, ensuring a high-quality set of variants. 

Our data is composed of 2,586 patients (samples), where each patient has a variable number of breaks (features). In this dataset there are 21 possible types of cancer (\eg breast cancer, liver cancer, pancreas cancer, \etc). At the same time, every sample comes from one of four possible germ layers (\ie basic cell types). The germ layers are \textit{ECTODERM}, \textit{ENDODERM}, \textit{MESODERM} and \textit{NEURAL CREST}. In this study we consider 4 types of chromosomal breaks: deletions (\textit{DEL}), translocations (\textit{TRA}), duplications (\textit{DUP}) and two kinds of inversions (\textit{t2tINV} and \textit{h2hINV}).
 
The total length of the human genome is over three billion base pairs. Those pairs are divided into 24 chromosomes. Characterizing breaks at a base pair level would end up with very sparse data on a high dimensional space (2,586 samples over 3 billion pairs). For this reason, we choose to reduce the granularity of the features used, working instead at chromosome level (\eg instead of considering a deletion on the gene 15p5, we consider a deletion on chromosome 15). As a result, we end up with 2,586 samples over the 24 chromosomes.
 
We use the germ layers as a generalization of the cancer types because of the small number of data samples, but with more data, this methodology could extract the markers for every cancer type. Table \ref{tab:histologies} shows how many samples are in the dataset for each cancer type, and how are these samples distributed over the germ layers.

\begin{table}[t]
\rowcolors{1}{gray!25}{white}
\begin{tabular}{lcccc}
\toprule
\rowcolor{white}
 &  ECTODERM &  ENDODERM &  MESODERM &  NEURAL\_CREST \\
\midrule
Biliary         &         0 &        34 &         0 &             0 \\
Bladder         &         0 &        23 &         0 &             0 \\
Bone/SoftTissue &         0 &         0 &        92 &             0 \\
Breast          &       209 &         0 &         0 &             0 \\
CNS             &         0 &         0 &         0 &           261 \\
Cervix          &         0 &         0 &        20 &             0 \\
Colon/Rectum    &         0 &        60 &         0 &             0 \\
Esophagus       &         0 &        87 &         0 &             0 \\
Head/Neck       &         0 &         0 &        56 &             0 \\
Kidney          &         0 &         0 &       176 &             0 \\
Liver           &         0 &       322 &         0 &             0 \\
Lung            &         0 &        84 &         0 &             0 \\
Lymphoid        &         0 &         0 &       197 &             0 \\
Myeloid         &         0 &         0 &        29 &             0 \\
Ovary           &         0 &         0 &       112 &             0 \\
Pancreas        &         0 &       306 &         0 &             0 \\
Prostate        &         0 &       263 &         0 &             0 \\
Skin            &         0 &         0 &         0 &           106 \\
Stomach         &         0 &        72 &         0 &             0 \\
Thyroid         &         0 &        30 &         0 &             0 \\
Uterus          &         0 &         0 &        47 &             0 \\ \midrule
Total           &       209 &      1281 &       729 &           367 \\
\bottomrule
\end{tabular}
\caption{Number of samples for every cancer type on the dataset, with their corresponding germ layer. \label{tab:histologies}}
\end{table}


\subsection{Preprocessing} \label{sec:preprocessing}

The first pre-processing step performed on the data was to remove all the patients with no germ layers nor cancer type labels available. After removing these unlabelled samples, we extracted a set of features from additional metadata available in the dataset (see the bottom part of Table \ref{tab:features}) and engineered another set (see the top part of Table \ref{tab:features}). The engineered features are related to the kind of breaks or their position in the DNA. 
Features include, among others, the number of rearrangements on every chromosome or the number of rearrangements of each type. 

Metadata include gender, age, tumour\_stage1 and tumour\_stage2 as features and germ layer as target variable (i.e., label). The last two features stand for the clinical stage of the tumour (non-genetic information). We initially had two options to select as target variable: the cancer type and the germ layer. Ideally, we would have selected cancer type but, the number of samples (2,586) \wrt the high number of cancer types (21) makes the problem unfeasible. For this reason, we end up selecting the germ layer instead of as a generalization of the cancer type. The final set of features (engineered and metadata features) as shown in Table \ref{tab:features} accompanied by a brief description.

\begin{table}[t]
\rowcolors{1}{gray!25}{white}
\begin{tabular}{ll|r}
\rowcolor{white}
\rowcolor{white}
\textbf{Genetic Data}                      &                                         & \textbf{Num. Features}\\ 
\midrule
\#\_of\_breaks                             &   No. of breaks of the sample.          & 1   \\
DUP, DEL, TRA                              &   No. of breaks per break type.         & 3  \\
h2hINV, t2tINV                             &   No. of breaks per inversion type.     & 2  \\
chr\_1, ..., chr\_Y                        &   No. of breaks per chromosome.         & 24  \\
DEL\_1, ..., DEL\_Y                        &   No. of deletions per chromosome.      & 24  \\
DUP\_1, ..., DUP\_Y                        &   No. of duplications per chromosome.   & 24  \\
TRA\_1, ..., TRA\_Y                        &  No. of translocations per chromosome.  & 24  \\
h2hINV\_1, ..., h2hINV\_Y                  &  No. of h2h inversions per chromosome.  & 24  \\
t2tINV\_1, ..., t2tINV\_Y                  &  No. of t2t inversions per chromosome.  & 24  \\ 
prop\_\{chr\_n, ..., t2tINV\_n\}           &   For each break type, for each chromosome,    & 149  \\ 
\rowcolor{gray!25}

& proportion of breaks over total breaks in patient. & \\ 
\midrule
\rowcolor{white}
\textbf{Patient Metadata}           &                                         &     \\ \midrule
female                                     & Gender of the patient, 1 if female and 0 otherwise.                  & 1   \\
donor\_age                                        & Age of the patient.                     & 1   \\
ts\_1\_category                           & Methastasic, Primary or Recurrent           & 3 \\

ts\_2\_category & NOS, bone marrow, periphleal blood, derived from  & 9 \\ 
\rowcolor{gray!25}
&  tumour, methastasis to lymph node, methastasis  &  \\
                &  to distant location, other or solid tissue & \\ \midrule
\textbf{Total number of features} & & 313 \\
\bottomrule
\end{tabular}
\caption{Genetic features (top) and metadata features (bottom) extracted from the dataset. \label{tab:features}}
\end{table}

After the extraction of these features, we perform a one-hot-encoding over all categorical features (both tumour stages). Furthermore, we impute 119 missing values for the age feature using the Multivariate Imputation by Chained Equations (MICE) \cite{buuren2010mice}.

Data was split into two partitions. One for training the model and another one for testing the classification results. The partition was stratified \wrt the germ layers, in order to try to maintain their original distribution (Table \ref{tab:histologies}). This way, we obtained a training partition with 2,068 samples and a test partition with 519 samples. Both of them containing 313 features, including boolean features from the one-hot-encoding.

\section{Approach}\label{sec:approach}
On this study, a Random Forest model \cite{scikit-learn} was used to identify which chromosomal rearrangements, and in which location, contain genetic markers. Random Forest has shown to have good performance over many applications, is one of the more interpretable models among the current machine learning state of the art, and it is capable of providing feature importance after training it. These properties make it a \textit{good} candidate for the computational biology field, both as a classification or feature selection tool \cite{random_forest_for_bioinformatics,random_forest_gene_selection}.  

We trained the Random Forest using the set of 313 features extracted from the chromosomal rearrangement data and the patient metadata (see Table \ref{tab:features}). The result of training this model is the feature importance order (where the best features are positioned at the beginning, and the worse are positioned at the end) and the classification results.

In the first experiment,  the feature importance orders obtained from training 300 Random Forest were used, each one containing 100 Decision Tree classifiers, to generate our aggregated feature ranking. This large number provides additional robustness to the aggregated feature ranking. The aggregated ranking was used to extract the \textit{most} important features. These will be our potential genetic markers.

In the second experiment,  this process was repeated four times, discriminating each germ layer type against all the rest (\ie one vs all). Four new rankings representing the best features (\ie potential genetic markers) for each one of the specific germ layers were obtained.

\section{Experiments}\label{sec:experiments}
All Random Forest models target to classify the patients by the germ layer associated with their cancer. This classification has two outputs of interest: the feature importance and classification results. The feature importance order extracts the relationships between the features and the germ layers, while classification results prove that these results are relevant.

To estimate the hyperparameters of the Random Forest,  Random Search Cross-Validation \cite{random_cv} was used, with three validation partitions and the parameter distributions presented in Table \ref{tab:hyperparameters}. Those hyperparameters were tuned for the four class classification task (\ie all vs all), and fix their value on all further experiments.

    \begin{table}[t]
\rowcolors{1}{gray!25}{white}
\centering
\begin{tabular}{lcc}
\toprule
\rowcolor{white}
\textbf{Hyperparameter} & \textbf{Distribution} & \textbf{Best Value} \\
\midrule
max\_depth              & $unif(2,20)$                                                  & 13                  \\
min\_samples\_split     & $unif(2,11)$                                                  & 5                   \\
min\_samples\_leaf      & $unif(1,20)$                                                  & 3                   \\
bootstrap               & \textit{unif}([\textit{True}, \textit{False}])                & \textit{True}     \\
criterion               & \textit{unif}([\textit{gini}, \textit{entropy}])              &  \textit{entropy} \\
max\_features           & \textit{unif}([\textit{auto}, \textit{log2}, \textit{None}])  &  \textit{None}    \\
\midrule
class\_weight           & -                                                             & \textit{balanced} \\
n\_estimators           & -                                                             & 100               \\
\bottomrule
\end{tabular}
\caption{Distributions used on the random search cross-validation and the best hyperparameters selected for the Random Forest.} \label{tab:hyperparameters}
\end{table}

\subsection{Feature ranking generation}\label{sec:feature_ranking_extraction}

Since Random Forest feature selection has a certain level of stochasticity, we first assess the stability of the method by performing 300 independent runs. We compute the mean ranking of all features as an indicator of robustness. Results (shown in Table \ref{tab:feature_ranking}) indicate a very strong consistency among runs, which speaks for the relevance of all further experiments.

\begin{table}[t]
\rowcolors{1}{gray!25}{white}
\centering
\begin{tabular}{ccl}
\toprule
\rowcolor{white}
\textbf{Ranking} & \textbf{Mean position}           &  \textbf{Features} \\
\midrule
             1 &          1.000 &       donor\_age\_at\_diagnosis \\
             2 &          2.000 &                       female \\
             3 &          3.000 &  tumour\_stage1\_Primary\_tumour \\
             4 &          4.018 &    tumour\_stage2\_solid\_tissue \\
             5 &          4.992 &                          DEL \\
             6 &          7.200 &           tumour\_stage2\_other \\
             7 &          7.478 &                        chr\_8 \\
             8 &          7.500 &                          TRA \\
             9 &          7.914 &             number\_of\_breaks \\
            10 &         10.518 &               proportion\_DUP \\
            11 &         10.910 &               proportion\_DEL \\
            12 &         12.448 &      tumour\_stage2\_lymph\_node \\
            13 &         12.728 &             proportion\_chr\_9 \\
            14 &         15.136 &                       t2tINV \\
            15 &         16.184 &            proportion\_DEL\_14 \\
\bottomrule
\end{tabular}
\caption{The 15 \textit{most} characteristic features found by the 300 Random Forest runs. The second column contains the mean position of each feature over the 300 executions.} \label{tab:feature_ranking}
\end{table}

\subsection{Germ layer specific ranking generation}
To obtain a feature ranking specific for each germ cell, the Random Forest was trained to discriminate every germ layer from the other three, transforming the original multi-class problem with four germ layers into four binary problems classifying the target germ layer vs the other three (\eg \textit{ENDODERM} vs \textit{NON-ENDODERM}).

After transforming the data labels, we train 300 Random Forests for each germ layer and aggregate these rankings (see the details in Section \ref{sec:feature_ranking_extraction}). This way, we obtain four rankings, with the \textit{most} discriminating features for each one of the germ layers.

The results obtained (see Table \ref{tab:feature_ranking_all}) show different feature rankings for every classification experiment, especially on the chromosome related features. These results suggest that the presence or the type of rearrangements on specific chromosomes are related with the germ layer of the cancer cell. 

\begin{table}[t]
\centering
\rowcolors{1}{gray!25}{white}
\begin{tabular}{llllll}
\toprule
\rowcolor{white}
\textbf{Rank}&\textbf{All Germ}& \textbf{ECTODERM} &\textbf{NEURAL\_CRES}T &        \textbf{MESODERM} &      \textbf{ENDODERM} \\
\midrule
1  &         donor\_age &            female &         donor\_age &         donor\_age &         donor\_age \\
2  &            female &         donor\_age &       ts1\_Primary &         ts2\_blood &            female \\
3  &       ts1\_Primary &  ts2\_solid\_tissue &          prop\_DUP &    ts2\_lymph\_node &               DEL \\
4  &  ts2\_solid\_tissue &               TRA &            chr\_21 &               DEL &  ts2\_solid\_tissue \\
5  &               DEL &             chr\_8 &        prop\_chr\_9 &        prop\_TRA\_5 &       \#\_of\_breaks \\
6  &         ts2\_other &    prop\_h2hINV\_19 &             chr\_5 &            female &               TRA \\
7  &             chr\_8 &            TRA\_17 &        prop\_chr\_2 &       \#\_of\_breaks &    ts2\_lymph\_node \\
8  &               TRA &        prop\_DEL\_4 &       prop\_DUP\_12 &            chr\_19 &         ts2\_other \\
9  &       \#\_of\_breaks &            t2tINV &  ts2\_solid\_tissue &        prop\_chr\_3 &       ts1\_Primary \\
10 &          prop\_DUP &       prop\_TRA\_17 &             chr\_6 &       ts1\_Primary &         ts2\_blood \\
11 &          prop\_DEL &          prop\_DEL &        prop\_chr\_5 &        prop\_chr\_9 &        prop\_TRA\_5 \\
12 &    ts2\_lymph\_node &        prop\_chr\_9 &    ts2\_lymph\_node &       prop\_t2tINV &          prop\_DEL \\
13 &        prop\_chr\_9 &        prop\_chr\_5 &             chr\_1 &  ts2\_solid\_tissue &        prop\_chr\_1 \\
14 &            t2tINV &        prop\_chr\_4 &             DEL\_1 &    ts1\_Metastatic &          prop\_DUP \\
15 &       prop\_DEL\_14 &        prop\_TRA\_9 &       prop\_chr\_21 &          prop\_TRA &        prop\_chr\_4 \\
\bottomrule
\end{tabular}
\caption{Best 15 features found for each classification experiment. The second column (All Germ) shows the best features for the all vs all classification task; this column corresponds with the results on Table \ref{tab:feature_ranking}. Third to the sixth column shows the best features for the one vs all classification task. \label{tab:feature_ranking_all}}
\end{table}

\subsection{Classification results}


To have some intuition about the importance of the features extracted by the model and their relation with the data, the model was tested using different sets of the best features of the ranking. We report classification results using the best 5, 15, 25, 50, 100, and all 313 features based on its importance ranking (reported in Table \ref{tab:classification_results}). Best F1 measure is obtained by using the best 25 features in most of the experiments.



\begin{table}[t]
\centering
\rowcolors{1}{gray!25}{white}
\begin{tabular}{c|c|cccc}
\toprule
\rowcolor{white}
 {}           & \textbf{F1 All Germ} &  \textbf{F1-ECTO.} &  \textbf{F1-NEURAL.} &  \textbf{F1-MESO.} &  \textbf{F1-ENDO.} \\
\midrule
\textbf{All } &     0.694 &       0.420 &           0.877 &       0.649 &       0.830 \\
\textbf{ 100} &     0.709 &       0.494 &           0.839 &       0.664 &       0.840 \\
\textbf{ 50 } &     0.722 &       0.543 &           0.859 &       0.660 &       0.826 \\
\textbf{ 25 } &  \textbf{0.741} &       \textbf{0.556} &           \textbf{0.887} &       0.681 &       \textbf{0.841} \\
\textbf{ 15 } &     0.737 &       0.551 &           0.879 &       \textbf{0.682} &       0.834 \\
\textbf{ 5  } &     0.630 &       0.404 &           0.818 &       0.565 &       0.731 \\
\bottomrule
\end{tabular}
\caption{Classification results using the top 5, 15, 50 and 100 features, or all of them. The second column (F1 All Germ) shows the mean F1 measure for the all vs all classification task. Third, to the sixth column show the F1 measure for the one vs all classification task. Best results in bold.\label{tab:classification_results}}
\end{table}

\section{Query-based evaluation}\label{sec:QBE}

The validation of the results obtained in the previous sections is not straight-forward. It will require thorough analysis from medical experts, in order to validate the existence of genetic markers associated with the identified features. This process can take from months to years.

In order to produce the first evaluation of the produced results, we use a crowd-based approach based on state of the art on cancer research. The well-known Medical Subject Headings \cite{mesh} was used, which indexes medical papers from PubMed \cite{pubmed}. PubMed allows querying over 29 million medical abstracts from MEDLINE, life science journals, and online books. Through its search engine, it is possible to find the number of papers mentioning both a cancer type (\eg Pancreas) and a chromosome. The result of this search can give  an idea of the current medical knowledge regarding the relationship between a type of cancer and a chromosome.

The obtained features are trained to discriminate germ layers. However, most medical papers do not work at this granularity. Instead, cancer type queries produce larger, and thus more representative results. For this reason we can only use this evaluation method reliably on the ECTODERM germ type, which only contains one cancer type (breast). The other germ types contain several cancer types on variable proportions. 
On the other hand, the engineered features are not appropriate for a straight-forward query on medical papers. These are often too specific (\eg proportion of h2h inversions on chromosome 19). For this reason, we limit ourselves to evaluate the relation found with whole chromosomes, disregarding any further particularity of the feature (\eg chromosome 19, instead of the proportion of h2h inversions on chromosome 19).

In particular, 24 queries were performed. One for each chromosome, together with the term for \textit{breast cancer}. Since not all chromosomes are expected to be related to breast cancer, not all queries will be relevant. We focus on the top three chromosomes related to breast cancer by the number of returned results. These are chromosomes 17, 11 and 8. At the same time, we find the top three chromosomes associated with a feature discriminating breast cancer on our results. These are chromosomes 8, 19 and 17 (chr\_8, prop\_h2hINV\_19 and TRA\_17). 

Remarkably, two of the three mentioned chromosomes on papers related to breast cancer, are involved in two of the three most relevant features we found for discriminating the ECTODERM germ layer (\ie breast cancer). This combination has a random statistical probability of roughly 1\%. 

\section{Discussion}

The results that are shown in this work open up several questions. To start with, the \textit{query-based evaluation} finds two of the top three chromosomes related to breast cancer to be consistent with the literature (see \S\ref{sec:QBE}). However, what is happening with those missing?

The chromosome found by the model, but not in the literature, is chromosome 19. This could be either a mistake by the model or an  unexplored relation  by the medical community. This is precisely the sort of relation with potential impact in the medical domain, as previously unknown genetic markers could be contained in this chromosome.

The chromosome found in the literature, but not by the model, is chromosome 11. This behaviour can be a consequence of its relevance for certain MESODERM cancer types (Kidney, where it is the first chromosome in several papers, or Ovarian, where it is the second). This, in turn, affects the Random Forest model, since this chromosome will not be discriminative for ECTODERM, even though it may be representative.

The absence of chromosome 11 in our results brought to our attention a couple of limitations on our approach. The first is related to the use of a classifier for extracting feature information. A classifier focuses on the discriminability of features. This might cause the model to oversee features that, while being representative for a particular type of cancer, are not discriminant in the context of several cancer types. This problem might be mitigated by doing pairwise classification instead of one vs all, comparing pairs of cancer types. By doing so, all features that are discriminant for our target cancer type with regards to any of the other cancer types would be identified. 

The complete, and unfeasible, solution to this problem would be to have a healthy person sample to compare against. However, healthy genomes do not have chromosomal rearrangements. As such, a healthy sample would be empty and impossible to compare against.

\section{Conclusions}


The results presented in this paper target the guidance of genetic markers. Given the large granularity of features used in our approach, this is not a straight-forward process from the medical perspective. To provide some evidence on the consistency of our approach, we performed a partial, query-based validation against an extensive database of medical papers. In this evaluation, we found that, out of the top three chromosomes identified with breast cancer in the literature, two are also found by the method. This coincidence has a random statistical probability of roughly 1\%. This gives us proportional confidence in asserting that the features found by our models are useful guidelines for cancer genetic markers. There is other evidence highlighting the medical consistency of our findings. For example, the most reliable feature for discriminating ECTODERM (\ie breast cancer) is gender. 

Another interesting insight from Table \ref{tab:feature_ranking_all} is that different germ layers seem to be related to different break types. While translocations are the most relevant break type for ECTODERM, for MESODERM and ENDODERM deletions seem to be more relevant. The NEURAL CREST case deserves a specific commentary. This cancer type is frequently related to children, in particular, Central Nervous System cancer (CNS). Children develop cancer differently when compared to adults. Our intuition is that chromosomal duplication is more consistent with the growth patterns of children, which would explain the findings of our model.

The same Table \ref{tab:feature_ranking_all} also displays a remarkable correlation between the specificity of a germ layer (\ie how many different cancer types it contains) and the specificity of the features found by our model. On the one hand, the most specific germ layers (ECTODERM and NEURAL CREST, with only one and two cancer types) have between 5 and 6 chromosome specific features among the top 10 ranked. On the other, the most generic germ layers (ENDODERM and MESODERM) and the all vs all classification (All Germ) have between zero and two chromosome specific features on the top 10. This seems to indicate that specific cancer types could be characterized further if more detail became available for analysis.

Randomized Decision Trees build inside the Random Forest algorithm, are among the fastest machine learning models for classification, with a complexity of \textit{O(KNlogN)} \cite{louppe2014understanding}. An essential feature of our model and methodology is thus its high scalability. If more data becomes available, we could extract more specific markers for one or more cancer types with a minimal computational cost. 
Beyond being scalable, the method is also robust to high-dimensional and sparse domains, since we treated these appropriately. Notice the actual train data set has 313 features for 2,068 samples, with a 92\% of zero values. In this case, the model design was tuned explicitly for this setting, including a large number of decision trees on each random forest, and a large number of random forests to be aggregated.

Summarizing the results obtained on this paper; we have obtained potential general markers that could be related to tumour-genesis on the four basic germ layer types. We have found specific potential markers for every one of the significant germ layers, obtaining coherent results respect to the known bibliography on the subject. Finally, we have obtained a general method for genetic marker mining, that could be generalized by the growth of the dataset. The continuation of this work requires extensive experimentation by medical experts in order to test and validate our many hypotheses.



\section*{Acknowledgements}
This work is partially supported by the Joint Study Agreement no. W156463 under the IBM/BSC Deep Learning Center agreement, by the Spanish Government through Programa Severo Ochoa (SEV-2015-0493), by the Spanish Ministry of Science and Technology through TIN2015-65316-P project, and by the Generalitat de Catalunya (contracts 2017-SGR-1414). Pro. U. Cort\'es is a member of the Sistema Nacional de Investigadores (Level III) (SNI-III). M\'exico.
We would like to thank MD. Adri\'an Puche Gallego and PhD. Davide Cirillo for useful discussions and guidance.
\bibliographystyle{ieeetr}
\bibliography{biblio}

\end{document}